\documentclass[prl,twocolumn,showpacs,preprintnumbers,amsmath,amssymb,superscriptaddress]{revtex4}

\usepackage{graphicx}
\usepackage{dcolumn}
\usepackage{bm,color}
\usepackage{amsmath, latexsym, amssymb, amscd, amsfonts, amsthm, epsfig, mathrsfs, graphicx, url,verbatim}
\usepackage[english]{babel}
\usepackage[braket]{qcircuit}

\begin{document}

\title{Quantum simulating an experiment: \\ Light interference from single ions and their mirror images}

\author{Luc Bouten}
\author{G\'e Vissers}
\affiliation{Q1t BV, Lindenlaan 15, 6584 AC, Molenhoek, www.q1t.nl}

\author{Ferdinand Schmidt-Kaler}
\affiliation{QUANTUM, Institut f\"ur Physik, Johannes Gutenberg-Universit\"at Mainz, Staudingerweg 7,  55128 Mainz, Germany}

\date{\today}
\pacs{02.60.Cb, 03.65.Yz, 03.67.-a, 03.67.Lx, 12.20.-m, 42.50.-p, 42.50.Lc, 42.50.Pq}

\begin{abstract}
We widen the range of applications for quantum computing by introducing digital
quantum simulation methods for coherent light-matter interactions: We simulate
an experiment where the emitted light from a single ion was interfering with its
mirror image [Eschner et. al., Nature 431, 495 (2001)]. Using the quantum
simulation software \texttt{q1tsim} we accurately reproduce the interference
pattern which had been observed experimentally and also show the effect of the
mirror position on the spontaneous emission rate of the ion. In order to minimize
the number of required qubits we implement a qubit-reinitialization technique.
We show that a digital quantum simulation of complex experiments in atomic and
quantum physics is feasible with no more than seven qubits, a setting which is
well within reach for advanced quantum computing platforms.
\end{abstract}

\maketitle

Typical applications of quantum simulation include open questions in solid state
physics \cite{Trabesinger12,Schaetz13,Georgescu14}. More recently, using trapped
ion setups, high energy physics problems have been addressed in experimental quantum
simulation \cite{Gerritsma11,Martinez16,Muschik_2017} and in hybrid classical-quantum
simulations solutions to molecular chemistry calculations have been
demonstrated \cite{Hempel18,Nam19}. Yet another set of applications investigates
energy transport in the quantum regime, with implications for our understanding
of biological systems \cite{Dylan18,Debnath18,Maier19}. Experimental realizations
of open quantum systems require the ability to implement both coherent many-body
dynamics and dissipative processes \cite{Mueller_2011,Schindler13}
Quantum simulation has been proposed even for mimicking non-physical
systems \cite{Lee15}. Using a superconducting circuit quantum computer problems
in the financial sector have been addressed, e.g. for an analyis of market stability
or for pricing financial derivatives \cite{Orus19,Martin19}.
Here, we widen the spectrum of applications for digital quantum simulation further
and propose to digitally simulate an experimental outcome. Specifically, we are able
to accurately reproduce by quantum simulation the outcome of a trapped single ion
experiment, where an interference pattern has been observed experimentally \cite{ERSB01}.
We show in the  digital quantum simulation -- similar to the experiment -- the
strong influence of the mirror position on the spontaneous emission rate of the ion.
Implementing an experimental setting which includes the coherent emission and
absorption of a quantum light field, the optical excitation of two-level systems,
the interference of light and backaction on the atomic electronic levels, we
provide a wide and universal set of tools for digital simulation which may
therefore be applied to predict results in many other atomic, ionic, molecular and optical
experiments.

For simulating light-matter interactions in a digital quantum simulator, we
divide the electromagnetic field into spatial slices, each containing either zero or one
photon, or any coherent superposition of these states. In this way, the field
is modelled by a tensor network of qubits. At the points where the
electromagnetic field interacts with matter, e.g. a single atom, ion, or a
collection of atomic emitters, we introduce a unitary interaction matrix that
couples the field slice at that position with the matter system. This unitary
interaction represents one time step of the simulation. As the electromagnetic
field is propagating with the speed of light, qubits in the tensor network move
to the next field slice in every new time step of the computation. The tensor
network might contain loops, which means we can also model fully coherent feedback,
e.g. when back-reflecting the emitted light by semi-cavities.

\begin{figure}
\centering
\includegraphics[width=0.45\textwidth]{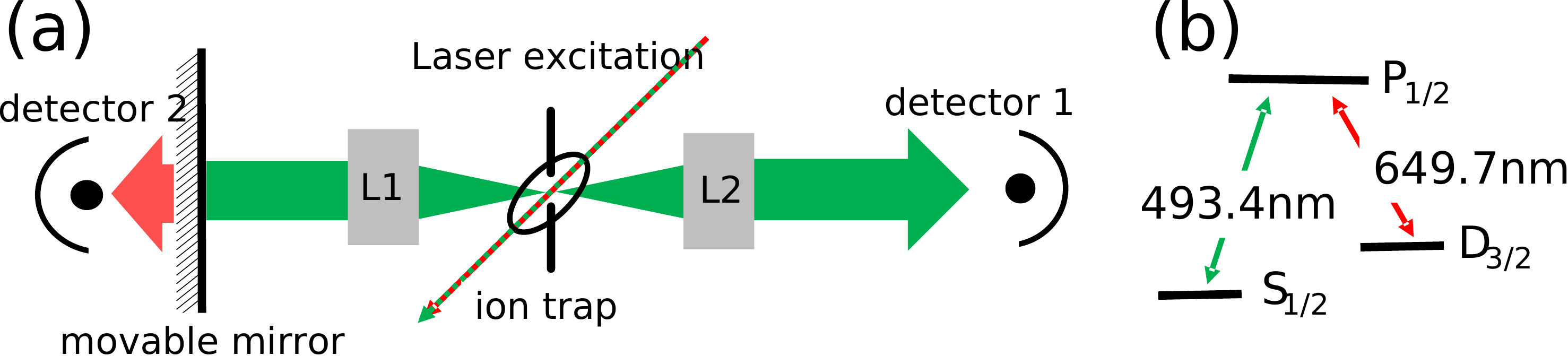}
\caption{\textbf{Setup:} (a) A single Barium$^+$ ion is trapped in a Paul trap.
The Ba$^+$ ion is driven on its $S_{1/2} \leftrightarrow P_{1/2}$
transition of 493~nm (green). The Ba$^+$ ion emits fluorescence photons
directly to a photo detector, via the focusing lens L2, and towards
a mirror reflecting 493~nm radiation. The light that travels towards the mirror is turned
into a collimated beam via the collimating lens L1. In this way two light paths
of different optical lengths towards the photo detector are created. The mirror
position can be actuated by a piezo electric stage and as such we can control the
difference in optical path length between the two light paths revealing an interference
pattern at the photo detector \cite{ERSB01}. The $P_{1/2} \leftrightarrow D_{3/2}$
transition near 650~nm is detected in detector 2, and allows for
revealing the population in $P_{1/2}$. (b) Scheme of relevant levels and transitions.}
\label{fig setup}
\end{figure}

The paper is organized as follows: After sketching the experimental setup to be
implemented by quantum simulation, we describe the model and its approximations.
We continue by a detailed discussion of the simulation calculation method and
exemplify the accuracy of results from the fitting parameters to the simulated
interference pattern, in comparison with the experimental findings.

As a case example of our open-system quantum simulation we model an experiment
\cite{ERSB01} in which a single ion is held
in a Paul trap in front of a mirror. When laser-exciting the ion, resonance
fluorescence is emitted, and two light paths towards a detector are established:
light that returns to the ion via the mirror before arriving at the detector, and
light directly being detected. If the optical path lengths of these two light
paths differ by a non-integer multiple of the transition wavelength
there will be destructive interference. By mounting the mirror on a piezo-electric
stage and varying the distance to the mirror, an interference pattern as a function
of the distance was observed.

In the experiment, a single Ba$^+$ ion is continuously laser-excited
and laser-cooled on its $S_{1/2} \leftrightarrow P_{1/2}$ and
$P_{1/2} \leftrightarrow D_{3/2}$ resonance lines
of 493~nm and 650~nm, respectively, see Fig.~\ref{fig setup}.

\textbf{The model:}
For the quantum simulation we ignore the $D_{3/2}$ state and model the Ba$^+$ ion
as a two-level system at a fixed position in space. The excitation of the transition
$S_{1/2}$ to $P_{1/2}$ induces Rabi oscillations as well as the emission of
fluorescence photons near 493~nm, which are subsequently collected by two lenses.
One lens collimates the light that is directed towards the mirror, such that light
can re-interact after a time delay and the second lens focuses the outgoing light
in the direction of the photo detector 1 \cite{ERSB01}, see Fig.~\ref{fig setup}(a).
The coordinates are fixed such that the Ba$^+$ ion is located at the origin and
we place the reflection at a position $-d$.
This leads to a natural time scale: we define $T$ as the time it
takes for a photon to make a round trip from the Ba$^+$ ion
to the mirror and back, i.e. $T := 2d/c$, where $c$ stands for the speed of light.
We divide the time interval $[0,T]$
in $N \in \mathbb{N}$ equal time slices and we define a discretization parameter
$\lambda$ by
\begin{equation}\label{eq definition lambda}
\lambda := \sqrt{\frac{T}{N}},
\end{equation}
such that every time slice represents $\lambda^2$ seconds.

\begin{figure}
\centering
\includegraphics[width=0.4\textwidth]{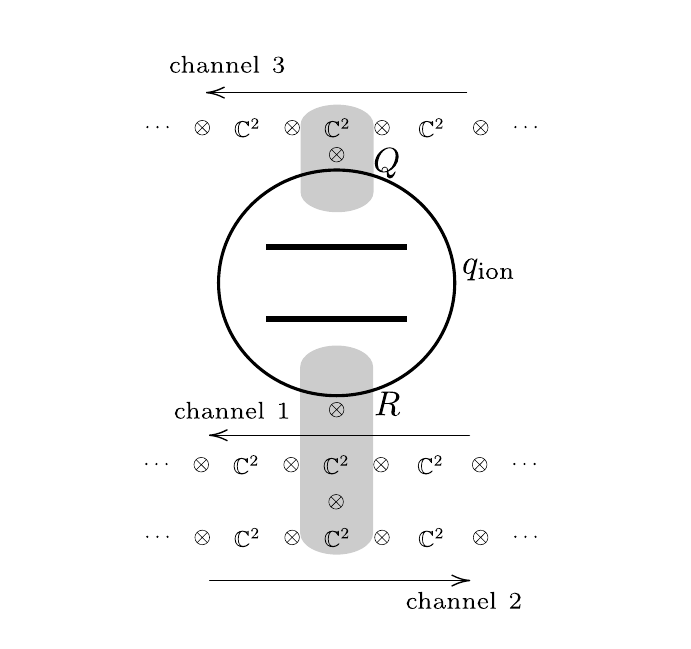}
\caption{\textbf{Model:} The ion is represented by the qubit at the center, labeled
$q_\mathrm{ion}$. Channel 1
represents the photons that are moving left towards the mirror. Channel 2 represents
the photons that are moving to the right towards the photo detector. Channel 3 is
the side channel for laser driving the ion.
Note that in the tensor network the side channel (channel 3) has also been represented
horizontally such that the network can be better visualized. Interactions $Q$ between
laser and the ion, and $R$ between the ion and the two photon field channels occur
at the origin.}
\label{fig repeated interaction horizontal}
\end{figure}

We divide the field, interacting with the single ion, into three channels:
(C1) Photons traveling from the ion to the mirror, (C2) Photons returning from
the mirror to the ion, and (C3) Laser light from the side exciting the ion.
All three channels are
represented by a doubly infinite string of qubits, see
Fig.~\ref{fig repeated interaction horizontal}.
The free time evolution $\Theta$ of the
electromagnetic field C1 and C3 corresponds to left shift, i.e. in one time step
all elements in the tensor network shift left by one position. For C2 a free time
evolution time step is imaged as a shift to the right, see
Fig.~\ref{fig repeated interaction horizontal}.

We now introduce an interaction
$R:\ \mathbb{C}^2 \otimes \mathbb{C}^2\otimes\mathbb{C}^2 \to
\mathbb{C}^2 \otimes \mathbb{C}^2\otimes\mathbb{C}^2$
between  the Ba$^+$ ion and the field slices at the origin of the channels 1 and 2 (last two copies of
$\mathbb{C}^2$ in the tensor product)
\begin{equation}\label{eq definition R}
R := e^{\sqrt{\kappa}\lambda(
\sigma_- \otimes \sigma_+ \otimes I  - \sigma_+\otimes \sigma_- \otimes I  +  \sigma_-\otimes I \otimes \sigma_+ -  \sigma_+ \otimes I \otimes \sigma_-)}.
\end{equation}
Here $\kappa$ is the strength of the coupling
between the Ba$^+$ ion and the two field channels. Without loss of generality we
assume identical coupling strength for C1 and C2, corresponding to an identical
focussing by lenses L1 and L2. Operators $\sigma_+$ and $\sigma_-$ denote
standard raising and lowering operators on a two-level system.

We introduce the interaction
$Q: \mathbb{C}^2 \otimes \mathbb{C}^2 \to \mathbb{C}^2 \otimes \mathbb{C}^2$
between the Ba$^+$ ion and the laser field slice at the origin of the
third channel
\begin{equation}\label{eq definition Q}
Q:= e^{\sqrt{\kappa_s}\lambda(\sigma_- \otimes \sigma_+  - \sigma_+ \otimes \sigma_- )},
\end{equation}
where $\kappa_s$ is the coupling strength between the Ba$^+$ ion
and the side channel C3.
Furthermore, the Ba$^+$ ion undergoes its own internal time evolution given by
\begin{equation}\label{eq definition L}
L :=  e^{ -i \omega \sigma_+\sigma_- \lambda^2 }.
\end{equation}

We initialize all field slices in C1 and C2 in the
vacuum state before interaction with the
ion. The side channel C3, however, is initialized
in a coherent state representing the resonantly driving laser.
A complex number $\alpha = |\alpha|e^{-i\omega l \lambda^2}$ represents its
amplitude and phase, where $l$ represents the time step. We now introduce
the discrete Weyl (or displacement) operator acting on
the qubit at the origin of C3
\begin{equation}\label{eq definition M}
M :=  e^{\lambda\alpha\sigma_+ - \lambda\overline{\alpha}\sigma_-}.
\end{equation}
Acting with the operator $M$ on the vacuum vector of the slice
of C3 at the origin, we drive this slice in a
coherent state that represents the resonant driving laser. In this way, Rabi oscillations are
induced in the ion with frequency $\Omega = |\alpha|\sqrt{\kappa_s}$.

Combining contributions from Eqns. \eqref{eq definition R} - \eqref{eq definition M},
we construct a time evolution which is given
by an evolution $U_l := (\Theta L RQM)^l$.
Repeated interactions as described by this evolution have been studied in
literature \cite{Kum85, Par88, LiP88, AtP06, BvH08, BHJ09, BGL18, ViB19, GoJ09}
and it can be shown that such a repeated interaction converges to a Hudson-Parthasarathy
quantum stochastic differential equation (QSDE) \cite{HuP84} in the limit where
the discretization parameter $\lambda$ goes to $0$.
QSDE's constitute the starting point for the
quantum stochastic input-output formalism introduced by Gardiner and Collett \cite{GC84}.
Consequently, our quantum simulation may be interpreted as a discretization of
input-output open quantum systems, optionally creating finite loops by connecting some
of the inputs to some of the outputs. In this specific case, this is the backreflection
of photons in C1 by the mirror to interact again with the ion as C2.

\begin{figure}
\centering
\includegraphics[width=0.4\textwidth]{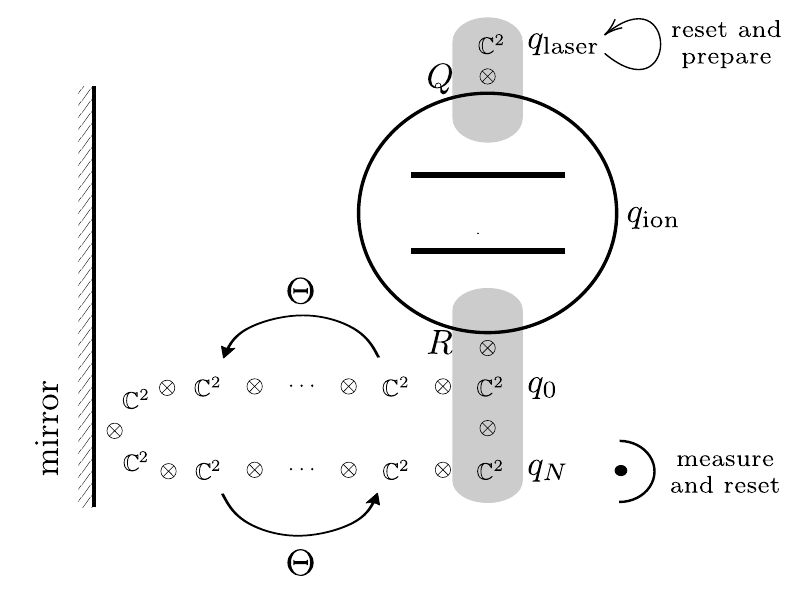}
\caption{\textbf{Closed loop interaction model:} Initially, the laser qubit $q_\mathrm{laser}$
interacts with the Ba$^+$ ion qubit $q_\mathrm{ion}$ through interaction $Q$.
Next, the ion interacts with the first and last field qubits $q_0$ and $q_N$
through interaction $R$. Afterwards, $q_\mathrm{laser}$ is reset, the outgoing
field qubit $q_N$ is measured and reset, and the field qubits are shifted.}
\label{fig closed loop}
\end{figure}

At the mirror, the field slices of C1 are transferred to
C2. In this way a loop of $N+1$ field qubits $q_0 \ldots q_N$ is created, see
Fig.~\ref{fig closed loop}. In the experiment \cite{ERSB01} the mirror is placed
at a distance of about 0.25~m, however modelling such a long time delay would require
a prohibitively large number of qubits in C1 and C2. Instead, we limit the distance $d$
to the wave length of the atomic transition and capture two full cycles of the interference
pattern. Note that by varying $d$ for given $N$, we also vary the time step
$\lambda$ according to Eqn.~\eqref{eq definition lambda}.

As soon as the last qubit $q_N$ in C2 has interacted with the ion we
project it in the $\sigma_z$ basis. If the measurement result is
$+1$ we  rotate the qubit back to $|0\rangle$. Then,
the qubit is shifted to C1 at the origin. In this way we reinitialize
the qubit and can re-use it in the quantum computation,
keeping the total required number of qubits minimal.
Employing a similar procedure for C3, we can simulate the entire channel with a
single qubit, see Fig.~\ref{fig closed loop}.

In the following we assume a 7 qubit quantum computer, such that one can model one
wavelength with 5 qubits and resolve expected sinusoidal interference pattern
sufficiently well. The 6th qubit represents the coherent laser driving and the
7th qubit the simplified two-level system of the ion.

\textbf{Quantum circuit:}\label{sec quantum circuit}
To implement the interaction described by the evolution $U_l$ all contributions
are mapped to elementary single- and two-qubit gate operations.
Leaving the interaction $R$ between the ion and the two photon field slices unspecified
for the moment, the circuit for time step $l$ is given by:
\begin{widetext}
\begin{equation}
\Qcircuit @C=0.7em @R=.5em {
    \lstick{q_0}         & \qw & \qw & \qw & \qw
                         & \qw & \gate{R} \qwx[4] & \qw & \qw
                         & \qw & \qswap \qwx[1] & \cds{4}{\cdots} & \qw
                         & \qswap \qwx[3] & \qswap \qwx[4] & \qw \\
    \lstick{q_1}         & \qw & \qw & \qw & \qw
                         & \qw & \qw & \qw & \qw
                         & \qw & \qswap & \qw & \qw
                         & \qw & \qw & \qw \\
    \lstick{\cdots}      & 
                           *+<1em,.9em>{\hphantom{X}} \\
    \lstick{q_{N-1}}     & \qw & \qw & \qw & \qw
                         & \qw & \qw & \qw & \qw
                         & \qw & \qw & \qw & \qw
                         & \qswap & \qw & \qw \\
    \lstick{q_N}         & \qw & \qw & \qw & \qw
                         & \qw & \multigate{1}{R} & \qw & \meter
                         & \targ & \qw & \qw & \qw
                         & \qw & \qswap & \qw \\
    \lstick{q_\mathrm{ion}}
                         & \qw & \qw & \ctrl{1} & \gate{R_y(-2\sqrt{\kappa_s}\lambda)}
                         & \ctrl{1} & \ghost{R} & \gate{R_z(-\omega \lambda^2)} & \qw
                         & \qw & \qw & \qw & \qw
                         & \qw & \qw & \qw \\
    \lstick{q_\mathrm{laser}}
                         & \gate{R_y(2|\alpha|\lambda)} & \gate{R_z(-\omega l\lambda^2)} & \targ & \ctrl{-1}
                         & \targ & \qw & \qw & \qw
                         & \qw & \push{~\ket{0}~} \ar @{|-{}} [0,-1] & \qw & \qw
                         & \qw & \qw & \qw \\
    \lstick{\mathrm{Detector}}
                         & \cw & \cw & \cw & \cw
                         & \cw & \cw & \cw & \cw \cwx[-3]
                         & \cctrl{-3} & \cw & \cw & \cw
                         & \cw & \cw & \cw
}
\label{circuit overall}
\end{equation}
\end{widetext}
Here the first two rotation gates represent the initialization of the laser.
Note that the $R_z(-\omega l\lambda^2)$ rotation setting the phase of
the laser qubit is different for each time step. The subsequent
three controlled gates simulate the interaction of the driving laser with the ion.
Next follows the
interaction of the ion with the photon field slices, and the internal evolution of the atom.
Finally the outgoing field slice is measured and reset, the laser qubit is reset,
and the field slices are shifted by a series of swap gates.

The interaction $R$ between the ion and the photon field, given by
Eqn.~\eqref{eq definition R}, can be decomposed into elementary quantum gates
as follows:
\begin{equation}
\Qcircuit @C=0.5em @R=.5em {
     \lstick{q_\mathrm{ion}}
        & \targ & \qw & \ctrl{2} & \gate{H}
        & \gate{S} & \ctrl{1} & \gate{\tilde{R}^\dagger} & \gate{\tilde{R}}
        & \ctrl{1} & \gate{S^\dagger} & \gate{H} & \ctrl{2}
        & \qw & \targ & \qw \\
    \lstick{q_0}
        & \qw & \targ & \qw & \ctrl{-1}
        & \targ & \targ & \ctrl{-1} & \qw
        & \targ & \targ & \ctrl{-1} & \qw
        & \targ & \qw & \qw \\
    \lstick{q_N}
        & \ctrl{-2} & \ctrl{-1} & \targ & \gate{V}
        & \ctrl{-1} & \qw & \qw & \ctrl{-2}
        & \qw & \ctrl{-1} & \gate{V^\dagger} & \targ
        & \ctrl{-1} & \ctrl{-2} & \qw
}
\label{circuit full interaction}
\end{equation}
where $\tilde{R} = R_y(2\sqrt{2\kappa}\lambda)$.




\begin{figure}
\centering
\includegraphics[width=0.5\textwidth]{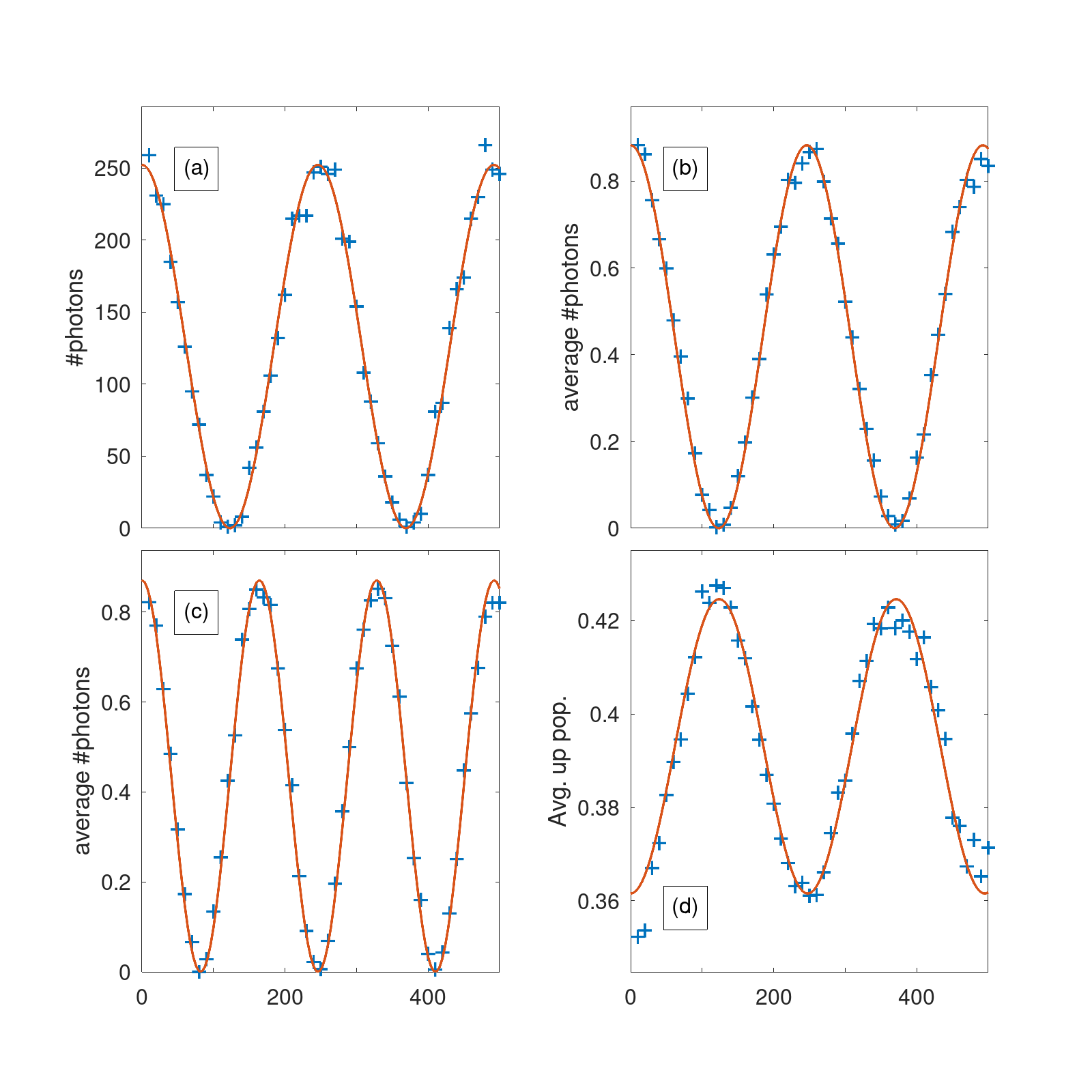}
\caption{\textbf{Simulation results:} (a) Calculated number of emitted photons after 25ps,
single run, with $\omega=1.0f$; (b) after 100fs, averaged over
1,000 runs, with $\omega=1.0f$; (d) after 100fs, averaged over 1,000 runs,
with $\omega=1.5f$; (d) Population of the upper state after 100 fs, time-averaged
and averaged over 1,000 runs. We use the same parameters as in panel (b).
Here, $f$ is the transition frequency of the Ba$^+$ ion (493nm). The other
parameters used in the calculation were $N=5$, $\Omega=0.01f$, $\kappa=6\times10^{12}$,
and $\kappa_s=3\times10^{13}$, for all panels. Calculated results from the
simulations are indicated by crosshair markers; solid lines show the fit. The fitted
wave lengths are 246.0 nm, 246.2 nm, 164.1 nm, and 247.5 nm for panels
(a)-(d) respectively.}
\label{fig pcount}
\end{figure}

\textbf{Results and discussion:}
We have implemented the quantum circuit from Eqn.~\eqref{circuit overall}
on a quantum computer simulator developed by Q1t BV, called \texttt{q1tsim} \cite{q1tsim}.

Two types of calculations have been performed. A direct simulation of the experiment was
done by running over a long time period (25 ps), but using only a single run.
The second type of calculation is more conducive to running on a real quantum computer,
since it consists of shorter runs simulating only 100 fs, but averaging the
results over multiple runs.

We set the transition frequency in the ion to the experimental value of
$f = 2\pi c / 493~\textrm{nm}$. To show that our approach may be experimentally
feasible to implement on current-technology quantum computers, we restricted the
calculations to using 5 qubits for representing the photon field.
The results for the simulations are shown in Fig.~\ref{fig pcount}(a)
and \ref{fig pcount}(b). Both simulations clearly show the interference
pattern that has been previously observed in the experiment \cite{ERSB01}.
Fitting the oscillation frequency of the
interference pattern, we find 246.0~nm and 246.2~nm, respectively, both close to the
expected value of 493/2~nm = 246.5~nm.

To prove that the observed pattern is indeed caused by interference, we have performed calculations with the value of the
internal transition frequency raised to $\omega=1.5f$. The results are shown
in Fig.~\ref{fig pcount}(c). As
expected, the fit for the wave length of the interference pattern of 164.1~nm
is shorter by the factor 1.5.

More interestingly, the interaction with the backreflected photon channel does not
only result in an interference pattern, but its back-action affects the emission
rate of the ion to be either enhanced or reduced, depending on the ion to mirror
distance. This results in a modulation of the $P_{1/2}$ state occupation probability,
which has been revealed from the observation of photons near 650~nm on the
$P_{1/2}$ to $D_{3/2}$ transition \cite{ERSB01}. The mirror coating in the
experimental realization was chosen highly transmitting near 650~nm such that
the detector 2 allows to detect such photons, see Fig.~\ref{fig setup}. To
illustrate that our simulation captures this behavior, the calculated population
of the upper state of the ion is evaluated, as a function of the distance, and
averaged over time and 1,000 runs, see Fig.~\ref{fig pcount}(d). It is easily
seen that the $P_{1/2}$ lifetime of the ion perfectly anticorrelates with the
emitted photon count rate at 493~nm.

Using the full circuit defined by Eqns \eqref{circuit overall} and \eqref{circuit full interaction},
a single time step in the simulation requires 7 single-qubit operations, $17 + N$
two-qubit operations, and two measurement operations, where SWAP gates and
controlled rotations are each counted as one two-qubit operation. The number
of time steps needed increases with the total simulated time. For the
results in Fig.~\ref{fig pcount}(b), between 120-600 steps have been taken for
each data point to simulate up to 100~fs, except for the shortest distances below 100~nm.
At $d=$10~nm, the required number of time steps runs up to 6,000 due to the small
value of $\lambda^2$. This may be mitigated though, by using a smaller number of
field qubits at these short distances. Furthermore, the interaction circuit
in Eqn.~\eqref{circuit full interaction} can be approximated using two separate
interactions between the ion and the field qubit, reducing the number of single
and double qubit gates to 3 and $6+N$ respectively.


\textbf{Conclusion and outlook:} We have been able to reproduce the interference pattern
which had been observed in experiment \cite{ERSB01} using the \texttt{q1tsim} simulator
of a quantum computer. Furthermore, we have shown that the presence of the mirror
modifies the emission by and thus the lifetime of an excited state of the Ba$^+$ ion.
Our results demonstrate simulation of a quantum model including an optical feedback
loop on a quantum computer. Using the methods we
presented in this paper, it will
be possible to model many more problems originating from
quantum optics, and more specifically cavity QED, on a future quantum computer.

Furthermore, we demonstrated that qubit re-initialization within a computation
run allows for reducing the number of required qubits, thus facilitating a
simulation of the experiment \cite{ERSB01} with $\leq$10 qubits. Circuits we have
been simulating on the \texttt{q1tsim} simulator are ready to be implemented
in the laboratory on state-of-the-art platforms.

\bibliography{Interference}

\begin{thebibliography}{29}
\expandafter\ifx\csname natexlab\endcsname\relax\def\natexlab#1{#1}\fi
\expandafter\ifx\csname bibnamefont\endcsname\relax
  \def\bibnamefont#1{#1}\fi
\expandafter\ifx\csname bibfnamefont\endcsname\relax
  \def\bibfnamefont#1{#1}\fi
\expandafter\ifx\csname citenamefont\endcsname\relax
  \def\citenamefont#1{#1}\fi
\expandafter\ifx\csname url\endcsname\relax
  \def\url#1{\texttt{#1}}\fi
\expandafter\ifx\csname urlprefix\endcsname\relax\def\urlprefix{URL }\fi
\providecommand{\bibinfo}[2]{#2}
\providecommand{\eprint}[2][]{\url{#2}}

\bibitem[{\citenamefont{Trabesinger}(2012)}]{Trabesinger12}
\bibinfo{author}{\bibfnamefont{A.}~\bibnamefont{Trabesinger}},
  \bibinfo{journal}{Nature Phys.} \textbf{\bibinfo{volume}{8}},
  \bibinfo{pages}{263} (\bibinfo{year}{2012}).

\bibitem[{\citenamefont{Schaetz et~al.}(2013)\citenamefont{Schaetz, Monroe, and
  Esslinger}}]{Schaetz13}
\bibinfo{author}{\bibfnamefont{T.}~\bibnamefont{Schaetz}},
  \bibinfo{author}{\bibfnamefont{C.~R.} \bibnamefont{Monroe}},
  \bibnamefont{and}
  \bibinfo{author}{\bibfnamefont{T.}~\bibnamefont{Esslinger}},
  \bibinfo{journal}{New Journal of Physics} \textbf{\bibinfo{volume}{15}},
  \bibinfo{pages}{085009} (\bibinfo{year}{2013}),
  \urlprefix\url{https://doi.org/10.1088%2F1367-2630%2F15%2F8%2F085009}.

\bibitem[{\citenamefont{Georgescu et~al.}(2014)\citenamefont{Georgescu, Ashhab,
  and Nori}}]{Georgescu14}
\bibinfo{author}{\bibfnamefont{I.~M.} \bibnamefont{Georgescu}},
  \bibinfo{author}{\bibfnamefont{S.}~\bibnamefont{Ashhab}}, \bibnamefont{and}
  \bibinfo{author}{\bibfnamefont{F.}~\bibnamefont{Nori}},
  \bibinfo{journal}{Rev. Mod. Phys.} \textbf{\bibinfo{volume}{86}},
  \bibinfo{pages}{153} (\bibinfo{year}{2014}).

\bibitem[{\citenamefont{Gerritsma et~al.}(2011)\citenamefont{Gerritsma, Lanyon,
  Kirchmair, Z\"ahringer, Hempel, Casanova, Garc\'{\i}a-Ripoll, Solano, Blatt,
  and Roos}}]{Gerritsma11}
\bibinfo{author}{\bibfnamefont{R.}~\bibnamefont{Gerritsma}},
  \bibinfo{author}{\bibfnamefont{B.~P.} \bibnamefont{Lanyon}},
  \bibinfo{author}{\bibfnamefont{G.}~\bibnamefont{Kirchmair}},
  \bibinfo{author}{\bibfnamefont{F.}~\bibnamefont{Z\"ahringer}},
  \bibinfo{author}{\bibfnamefont{C.}~\bibnamefont{Hempel}},
  \bibinfo{author}{\bibfnamefont{J.}~\bibnamefont{Casanova}},
  \bibinfo{author}{\bibfnamefont{J.~J.} \bibnamefont{Garc\'{\i}a-Ripoll}},
  \bibinfo{author}{\bibfnamefont{E.}~\bibnamefont{Solano}},
  \bibinfo{author}{\bibfnamefont{R.}~\bibnamefont{Blatt}}, \bibnamefont{and}
  \bibinfo{author}{\bibfnamefont{C.~F.} \bibnamefont{Roos}},
  \bibinfo{journal}{Phys. Rev. Lett.} \textbf{\bibinfo{volume}{106}},
  \bibinfo{pages}{060503} (\bibinfo{year}{2011}).

\bibitem[{\citenamefont{Martinez et~al.}(2016)\citenamefont{Martinez, Muschik,
  Schindler, Nigg, Erhard, Heyl, Hauke, Dalmonte, Monz, Zoller
  et~al.}}]{Martinez16}
\bibinfo{author}{\bibfnamefont{E.}~\bibnamefont{Martinez}},
  \bibinfo{author}{\bibfnamefont{C.}~\bibnamefont{Muschik}},
  \bibinfo{author}{\bibfnamefont{P.}~\bibnamefont{Schindler}},
  \bibinfo{author}{\bibfnamefont{D.}~\bibnamefont{Nigg}},
  \bibinfo{author}{\bibfnamefont{A.}~\bibnamefont{Erhard}},
  \bibinfo{author}{\bibfnamefont{M.}~\bibnamefont{Heyl}},
  \bibinfo{author}{\bibfnamefont{P.}~\bibnamefont{Hauke}},
  \bibinfo{author}{\bibfnamefont{M.}~\bibnamefont{Dalmonte}},
  \bibinfo{author}{\bibfnamefont{T.}~\bibnamefont{Monz}},
  \bibinfo{author}{\bibfnamefont{P.}~\bibnamefont{Zoller}},
  \bibnamefont{et~al.}, \bibinfo{journal}{Nature}
  \textbf{\bibinfo{volume}{534}}, \bibinfo{pages}{516} (\bibinfo{year}{2016}).

\bibitem[{\citenamefont{Muschik et~al.}(2017)\citenamefont{Muschik, Heyl,
  Martinez, Monz, Schindler, Vogell, Dalmonte, Hauke, Blatt, and
  Zoller}}]{Muschik_2017}
\bibinfo{author}{\bibfnamefont{C.}~\bibnamefont{Muschik}},
  \bibinfo{author}{\bibfnamefont{M.}~\bibnamefont{Heyl}},
  \bibinfo{author}{\bibfnamefont{E.}~\bibnamefont{Martinez}},
  \bibinfo{author}{\bibfnamefont{T.}~\bibnamefont{Monz}},
  \bibinfo{author}{\bibfnamefont{P.}~\bibnamefont{Schindler}},
  \bibinfo{author}{\bibfnamefont{B.}~\bibnamefont{Vogell}},
  \bibinfo{author}{\bibfnamefont{M.}~\bibnamefont{Dalmonte}},
  \bibinfo{author}{\bibfnamefont{P.}~\bibnamefont{Hauke}},
  \bibinfo{author}{\bibfnamefont{R.}~\bibnamefont{Blatt}}, \bibnamefont{and}
  \bibinfo{author}{\bibfnamefont{P.}~\bibnamefont{Zoller}},
  \bibinfo{journal}{New Journal of Physics} \textbf{\bibinfo{volume}{19}},
  \bibinfo{pages}{103020} (\bibinfo{year}{2017}).

\bibitem[{\citenamefont{Hempel et~al.}(2018)\citenamefont{Hempel, Maier,
  Romero, McClean, Monz, Shen, Jurcevic, Lanyon, Love, Babbush
  et~al.}}]{Hempel18}
\bibinfo{author}{\bibfnamefont{C.}~\bibnamefont{Hempel}},
  \bibinfo{author}{\bibfnamefont{C.}~\bibnamefont{Maier}},
  \bibinfo{author}{\bibfnamefont{J.}~\bibnamefont{Romero}},
  \bibinfo{author}{\bibfnamefont{J.}~\bibnamefont{McClean}},
  \bibinfo{author}{\bibfnamefont{T.}~\bibnamefont{Monz}},
  \bibinfo{author}{\bibfnamefont{H.}~\bibnamefont{Shen}},
  \bibinfo{author}{\bibfnamefont{P.}~\bibnamefont{Jurcevic}},
  \bibinfo{author}{\bibfnamefont{B.~P.} \bibnamefont{Lanyon}},
  \bibinfo{author}{\bibfnamefont{P.}~\bibnamefont{Love}},
  \bibinfo{author}{\bibfnamefont{R.}~\bibnamefont{Babbush}},
  \bibnamefont{et~al.}, \bibinfo{journal}{Phys. Rev. X}
  \textbf{\bibinfo{volume}{8}}, \bibinfo{pages}{031022} (\bibinfo{year}{2018}).

\bibitem[{\citenamefont{{Nam} et~al.}(2019)\citenamefont{{Nam}, {Chen},
  {Pisenti}, {Wright}, {Delaney}, {Maslov}, {Brown}, {Allen}, {Amini},
  {Apisdorf} et~al.}}]{Nam19}
\bibinfo{author}{\bibfnamefont{Y.}~\bibnamefont{{Nam}}},
  \bibinfo{author}{\bibfnamefont{J.-S.} \bibnamefont{{Chen}}},
  \bibinfo{author}{\bibfnamefont{N.~C.} \bibnamefont{{Pisenti}}},
  \bibinfo{author}{\bibfnamefont{K.}~\bibnamefont{{Wright}}},
  \bibinfo{author}{\bibfnamefont{C.}~\bibnamefont{{Delaney}}},
  \bibinfo{author}{\bibfnamefont{D.}~\bibnamefont{{Maslov}}},
  \bibinfo{author}{\bibfnamefont{K.~R.} \bibnamefont{{Brown}}},
  \bibinfo{author}{\bibfnamefont{S.}~\bibnamefont{{Allen}}},
  \bibinfo{author}{\bibfnamefont{J.~M.} \bibnamefont{{Amini}}},
  \bibinfo{author}{\bibfnamefont{J.}~\bibnamefont{{Apisdorf}}},
  \bibnamefont{et~al.}, \bibinfo{journal}{arXiv e-prints}
  \bibinfo{eid}{arXiv:1902.10171} (\bibinfo{year}{2019}), \eprint{1902.10171}.

\bibitem[{\citenamefont{Gorman et~al.}(2018)\citenamefont{Gorman, Hemmerling,
  Megidish, Moeller, Schindler, Sarovar, and Haeffner}}]{Dylan18}
\bibinfo{author}{\bibfnamefont{D.~J.} \bibnamefont{Gorman}},
  \bibinfo{author}{\bibfnamefont{B.}~\bibnamefont{Hemmerling}},
  \bibinfo{author}{\bibfnamefont{E.}~\bibnamefont{Megidish}},
  \bibinfo{author}{\bibfnamefont{S.~A.} \bibnamefont{Moeller}},
  \bibinfo{author}{\bibfnamefont{P.}~\bibnamefont{Schindler}},
  \bibinfo{author}{\bibfnamefont{M.}~\bibnamefont{Sarovar}}, \bibnamefont{and}
  \bibinfo{author}{\bibfnamefont{H.}~\bibnamefont{Haeffner}},
  \bibinfo{journal}{Phys. Rev. X} \textbf{\bibinfo{volume}{8}},
  \bibinfo{pages}{011038} (\bibinfo{year}{2018}).

\bibitem[{\citenamefont{Debnath et~al.}(2018)\citenamefont{Debnath, Linke,
  Wang, Figgatt, Landsman, Duan, and Monroe}}]{Debnath18}
\bibinfo{author}{\bibfnamefont{S.}~\bibnamefont{Debnath}},
  \bibinfo{author}{\bibfnamefont{N.~M.} \bibnamefont{Linke}},
  \bibinfo{author}{\bibfnamefont{S.-T.} \bibnamefont{Wang}},
  \bibinfo{author}{\bibfnamefont{C.}~\bibnamefont{Figgatt}},
  \bibinfo{author}{\bibfnamefont{K.~A.} \bibnamefont{Landsman}},
  \bibinfo{author}{\bibfnamefont{L.-M.} \bibnamefont{Duan}}, \bibnamefont{and}
  \bibinfo{author}{\bibfnamefont{C.}~\bibnamefont{Monroe}},
  \bibinfo{journal}{Phys. Rev. Lett.} \textbf{\bibinfo{volume}{120}},
  \bibinfo{pages}{073001} (\bibinfo{year}{2018}).

\bibitem[{\citenamefont{Maier et~al.}(2019)\citenamefont{Maier, Brydges,
  Jurcevic, Trautmann, Hempel, Lanyon, Hauke, Blatt, and Roos}}]{Maier19}
\bibinfo{author}{\bibfnamefont{C.}~\bibnamefont{Maier}},
  \bibinfo{author}{\bibfnamefont{T.}~\bibnamefont{Brydges}},
  \bibinfo{author}{\bibfnamefont{P.}~\bibnamefont{Jurcevic}},
  \bibinfo{author}{\bibfnamefont{N.}~\bibnamefont{Trautmann}},
  \bibinfo{author}{\bibfnamefont{C.}~\bibnamefont{Hempel}},
  \bibinfo{author}{\bibfnamefont{B.~P.} \bibnamefont{Lanyon}},
  \bibinfo{author}{\bibfnamefont{P.}~\bibnamefont{Hauke}},
  \bibinfo{author}{\bibfnamefont{R.}~\bibnamefont{Blatt}}, \bibnamefont{and}
  \bibinfo{author}{\bibfnamefont{C.~F.} \bibnamefont{Roos}},
  \bibinfo{journal}{Phys. Rev. Lett.} \textbf{\bibinfo{volume}{122}},
  \bibinfo{pages}{050501} (\bibinfo{year}{2019}).

\bibitem[{\citenamefont{Müller et~al.}(2011)\citenamefont{Müller, Hammerer,
  Zhou, Roos, and Zoller}}]{Mueller_2011}
\bibinfo{author}{\bibfnamefont{M.}~\bibnamefont{Müller}},
  \bibinfo{author}{\bibfnamefont{K.}~\bibnamefont{Hammerer}},
  \bibinfo{author}{\bibfnamefont{Y.~L.} \bibnamefont{Zhou}},
  \bibinfo{author}{\bibfnamefont{C.~F.} \bibnamefont{Roos}}, \bibnamefont{and}
  \bibinfo{author}{\bibfnamefont{P.}~\bibnamefont{Zoller}},
  \bibinfo{journal}{New Journal of Physics} \textbf{\bibinfo{volume}{13}},
  \bibinfo{pages}{085007} (\bibinfo{year}{2011}).

\bibitem[{\citenamefont{Schindler et~al.}(2013)\citenamefont{Schindler,
  M\"uller, Nigg, Barreiro, Martinez, Hennrich, Monz, Diehl, Zoller, and
  Blatt}}]{Schindler13}
\bibinfo{author}{\bibfnamefont{P.}~\bibnamefont{Schindler}},
  \bibinfo{author}{\bibfnamefont{M.}~\bibnamefont{M\"uller}},
  \bibinfo{author}{\bibfnamefont{D.}~\bibnamefont{Nigg}},
  \bibinfo{author}{\bibfnamefont{J.}~\bibnamefont{Barreiro}},
  \bibinfo{author}{\bibfnamefont{E.}~\bibnamefont{Martinez}},
  \bibinfo{author}{\bibfnamefont{M.}~\bibnamefont{Hennrich}},
  \bibinfo{author}{\bibfnamefont{T.}~\bibnamefont{Monz}},
  \bibinfo{author}{\bibfnamefont{S.}~\bibnamefont{Diehl}},
  \bibinfo{author}{\bibfnamefont{P.}~\bibnamefont{Zoller}}, \bibnamefont{and}
  \bibinfo{author}{\bibfnamefont{R.}~\bibnamefont{Blatt}},
  \bibinfo{journal}{Nature Physics} \textbf{\bibinfo{volume}{9}},
  \bibinfo{pages}{361} (\bibinfo{year}{2013}).

\bibitem[{\citenamefont{Lee et~al.}(2015)\citenamefont{Lee, Alvarez-Rodriguez,
  Cheng, Lamata, and Solano}}]{Lee15}
\bibinfo{author}{\bibfnamefont{T.~E.} \bibnamefont{Lee}},
  \bibinfo{author}{\bibfnamefont{U.}~\bibnamefont{Alvarez-Rodriguez}},
  \bibinfo{author}{\bibfnamefont{X.-H.} \bibnamefont{Cheng}},
  \bibinfo{author}{\bibfnamefont{L.}~\bibnamefont{Lamata}}, \bibnamefont{and}
  \bibinfo{author}{\bibfnamefont{E.}~\bibnamefont{Solano}},
  \bibinfo{journal}{Phys. Rev. A} \textbf{\bibinfo{volume}{92}},
  \bibinfo{pages}{032129} (\bibinfo{year}{2015}).

\bibitem[{\citenamefont{Orus et~al.}(2018)\citenamefont{Orus, Mugel, and
  Lizaso}}]{Orus19}
\bibinfo{author}{\bibfnamefont{R.}~\bibnamefont{Orus}},
  \bibinfo{author}{\bibfnamefont{S.}~\bibnamefont{Mugel}}, \bibnamefont{and}
  \bibinfo{author}{\bibfnamefont{E.}~\bibnamefont{Lizaso}},
  \bibinfo{journal}{arXiv:1810.07690}  (\bibinfo{year}{2018}).

\bibitem[{\citenamefont{{Martin} et~al.}(2019)\citenamefont{{Martin},
  {Candelas}, {Rodr{\'\i}guez-Rozas}, {Mart{\'\i}n-Guerrero}, {Chen}, {Lamata},
  {Or{\'u}s}, {Solano}, and {Sanz}}}]{Martin19}
\bibinfo{author}{\bibfnamefont{A.}~\bibnamefont{{Martin}}},
  \bibinfo{author}{\bibfnamefont{B.}~\bibnamefont{{Candelas}}},
  \bibinfo{author}{\bibfnamefont{{\'A}.}~\bibnamefont{{Rodr{\'\i}guez-Rozas}}},
  \bibinfo{author}{\bibfnamefont{J.~D.} \bibnamefont{{Mart{\'\i}n-Guerrero}}},
  \bibinfo{author}{\bibfnamefont{X.}~\bibnamefont{{Chen}}},
  \bibinfo{author}{\bibfnamefont{L.}~\bibnamefont{{Lamata}}},
  \bibinfo{author}{\bibfnamefont{R.}~\bibnamefont{{Or{\'u}s}}},
  \bibinfo{author}{\bibfnamefont{E.}~\bibnamefont{{Solano}}}, \bibnamefont{and}
  \bibinfo{author}{\bibfnamefont{M.}~\bibnamefont{{Sanz}}},
  \bibinfo{journal}{arXiv e-prints} \bibinfo{eid}{arXiv:1904.05803}
  (\bibinfo{year}{2019}), \eprint{1904.05803}.

\bibitem[{\citenamefont{Eschner et~al.}(2001)\citenamefont{Eschner, Raab,
  Schmidt-Kaler, and Blatt}}]{ERSB01}
\bibinfo{author}{\bibfnamefont{J.}~\bibnamefont{Eschner}},
  \bibinfo{author}{\bibfnamefont{C.}~\bibnamefont{Raab}},
  \bibinfo{author}{\bibfnamefont{F.}~\bibnamefont{Schmidt-Kaler}},
  \bibnamefont{and} \bibinfo{author}{\bibfnamefont{R.}~\bibnamefont{Blatt}},
  \bibinfo{journal}{Nature} \textbf{\bibinfo{volume}{413}},
  \bibinfo{pages}{495} (\bibinfo{year}{2001}).

\bibitem[{\citenamefont{K\"ummerer}(1985)}]{Kum85}
\bibinfo{author}{\bibfnamefont{B.}~\bibnamefont{K\"ummerer}},
  \bibinfo{journal}{J. Funct. Anal.} \textbf{\bibinfo{volume}{63}},
  \bibinfo{pages}{139} (\bibinfo{year}{1985}).

\bibitem[{\citenamefont{Parthasarathy}(1988)}]{Par88}
\bibinfo{author}{\bibfnamefont{K.~R.} \bibnamefont{Parthasarathy}},
  \bibinfo{journal}{Journal of applied probability}
  \textbf{\bibinfo{volume}{special volume 25A}}, \bibinfo{pages}{151}
  (\bibinfo{year}{1988}).

\bibitem[{\citenamefont{Lindsay and Parthasarathy}(1988)}]{LiP88}
\bibinfo{author}{\bibfnamefont{J.~M.} \bibnamefont{Lindsay}} \bibnamefont{and}
  \bibinfo{author}{\bibfnamefont{K.~R.} \bibnamefont{Parthasarathy}},
  \bibinfo{journal}{Sankhya: The Indian journal of statistics}
  \textbf{\bibinfo{volume}{50}}, \bibinfo{pages}{151} (\bibinfo{year}{1988}).

\bibitem[{\citenamefont{Attal and Pautrat}(2006)}]{AtP06}
\bibinfo{author}{\bibfnamefont{S.}~\bibnamefont{Attal}} \bibnamefont{and}
  \bibinfo{author}{\bibfnamefont{Y.}~\bibnamefont{Pautrat}},
  \bibinfo{journal}{Annales~Henri~Poincar\'e} \textbf{\bibinfo{volume}{7}},
  \bibinfo{pages}{59} (\bibinfo{year}{2006}).

\bibitem[{\citenamefont{Bouten and {van Handel}}(2008)}]{BvH08}
\bibinfo{author}{\bibfnamefont{L.~M.} \bibnamefont{Bouten}} \bibnamefont{and}
  \bibinfo{author}{\bibfnamefont{R.}~\bibnamefont{{van Handel}}},
  \bibinfo{journal}{J. Math. Phys.} \textbf{\bibinfo{volume}{49}},
  \bibinfo{pages}{102109} (\bibinfo{year}{2008}).

\bibitem[{\citenamefont{Bouten et~al.}(2009)\citenamefont{Bouten, {van Handel},
  and James}}]{BHJ09}
\bibinfo{author}{\bibfnamefont{L.~M.} \bibnamefont{Bouten}},
  \bibinfo{author}{\bibfnamefont{R.}~\bibnamefont{{van Handel}}},
  \bibnamefont{and} \bibinfo{author}{\bibfnamefont{M.}~\bibnamefont{James}},
  \bibinfo{journal}{SIAM Review} \textbf{\bibinfo{volume}{51}},
  \bibinfo{pages}{239} (\bibinfo{year}{2009}).

\bibitem[{\citenamefont{Belton et~al.}(2018)\citenamefont{Belton, Gnacik, and
  Lindsay}}]{BGL18}
\bibinfo{author}{\bibfnamefont{A.}~\bibnamefont{Belton}},
  \bibinfo{author}{\bibfnamefont{M.}~\bibnamefont{Gnacik}}, \bibnamefont{and}
  \bibinfo{author}{\bibfnamefont{J.}~\bibnamefont{Lindsay}},
  \bibinfo{journal}{Annales Henri Poincar\'e} \textbf{\bibinfo{volume}{19}},
  \bibinfo{pages}{1711} (\bibinfo{year}{2018}).

\bibitem[{\citenamefont{Vissers and Bouten}(2019)}]{ViB19}
\bibinfo{author}{\bibfnamefont{G.}~\bibnamefont{Vissers}} \bibnamefont{and}
  \bibinfo{author}{\bibfnamefont{L.}~\bibnamefont{Bouten}},
  \bibinfo{journal}{Quantum Information Processing}
  \textbf{\bibinfo{volume}{18}}, \bibinfo{pages}{152} (\bibinfo{year}{2019}),
  ISSN \bibinfo{issn}{1573-1332},
  \urlprefix\url{https://doi.org/10.1007/s11128-019-2272-z}.

\bibitem[{\citenamefont{Gough and James}(2009)}]{GoJ09}
\bibinfo{author}{\bibfnamefont{J.}~\bibnamefont{Gough}} \bibnamefont{and}
  \bibinfo{author}{\bibfnamefont{M.}~\bibnamefont{James}},
  \bibinfo{journal}{IEEE Trans. Aut. Control} \textbf{\bibinfo{volume}{54}},
  \bibinfo{pages}{2530} (\bibinfo{year}{2009}).

\bibitem[{\citenamefont{Hudson and Parthasarathy}(1984)}]{HuP84}
\bibinfo{author}{\bibfnamefont{R.~L.} \bibnamefont{Hudson}} \bibnamefont{and}
  \bibinfo{author}{\bibfnamefont{K.~R.} \bibnamefont{Parthasarathy}},
  \bibinfo{journal}{Commun. Math. Phys.} \textbf{\bibinfo{volume}{93}},
  \bibinfo{pages}{301} (\bibinfo{year}{1984}).

\bibitem[{\citenamefont{Gardiner and Collett}(1985)}]{GC84}
\bibinfo{author}{\bibfnamefont{C.~W.} \bibnamefont{Gardiner}} \bibnamefont{and}
  \bibinfo{author}{\bibfnamefont{M.~J.} \bibnamefont{Collett}},
  \bibinfo{journal}{Phys. Rev. A} \textbf{\bibinfo{volume}{31}},
  \bibinfo{pages}{3761} (\bibinfo{year}{1985}).

\bibitem[{\citenamefont{{Q1t BV}}(2019)}]{q1tsim}
\bibinfo{author}{\bibnamefont{{Q1t BV}}},
  \emph{\bibinfo{title}{{\texttt{q1tsim} quantum computer simulator}}},
  \bibinfo{howpublished}{\url{https://crates.io/crates/q1tsim},
  \url{https://github.com/Q1tBV/q1tsim}} (\bibinfo{year}{2019}),
  \bibinfo{note}{(version 0.3)}.

\end{thebibliography}

\end{document}